\documentclass[twocolumn,aps, amsfonts, pra]{revtex4}

\usepackage{epsfig}

\begin{document}

\title{
Family of Zeilinger-Horne-Greenberger ``W'' states leads to
stronger nonclassicality than family of Greenberger-Horne-Zeilinger ``GHZ''
states }

\author{Aditi Sen(De)\(^1\), Ujjwal Sen\(^1\), Marcin Wie{\' s}niak\(^2\), 
Dagomir Kaszlikowski\(^{3,4}\),
and Marek \.Zukowski\(^1\)}
\affiliation{
\(^1\)Instytut Fizyki Teoretycznej i Astrofizyki, Uniwersytet
Gda\'nski, PL-80-952 Gda\'nsk, Poland, \\
\(^2\)Wydzia{\l } Matematyki i Fizyki, Uniwersytet
Gda\'nski, PL-80-952 Gda\'nsk, Poland, \\
\(^3\)Department of Physics, National University of Singapore,
Singapore 117542, Singapore,\\
\(^4\)Instytut Fizyki Do{\' s}wiadczalnej, Uniwersytet
Gda\'nski, PL-80-952 Gda\'nsk, Poland
}

\begin{abstract}
The \(N\)-qubit states of the \(W\) class, for $N>10$, lead to more robust
(against noise admixture) violations of local realism, than the GHZ states.
These violations are most pronounced for
correlations for a pair of qubits, conditioned on  specific measurement results
for the remaining $N-2$ qubits.
The considerations provide us with a \emph{qualitative} difference between the \(W\) state and GHZ state
in the situation when they are separately sent via depolarizing channels. 
For sufficiently high amount of noise in the depolarizing channel, the GHZ states cannot produce a distillable state
between two qubits,  whereas the \(W\) states can still produce a 
distillable state in a similar situation. 

\end{abstract}
\maketitle

\newcommand{\tr}{{\rm tr}}

\section{Introduction}

The Greenberger-Horne-Zeilinger (GHZ) states
\cite{GHZ} give the maximal violation of correlation
function Bell inequalities \cite{WW, ZB}. The second best known
multiqubit states is the $W$ family
\cite{ZHG, CDV1}. Their
behavior is opposite in some respects to the GHZ family. For
example in the three qubit case, the \(W\) state  has the maximal
bipartite entanglement among all three qubit states \cite{CDV1},
whereas the GHZ state, has no bipartite entanglement (cf \cite{PopPle, CABELLO}).

In this paper,  we exhibit a kind of  complementarity
between the \(N\)-qubit \(W\) states  and GHZ states from the perspective
of robustness of the nonclassical correlations against white noise admixture
\cite{Acin1,Acin,CGLMP,Masanes}.
  For the \(N\)-qubit \(W\) state
\[
\left|W_N \right\rangle = \frac{1}{\sqrt{N}}\left(\left|100 \ldots 00\right\rangle +
\left|010 \ldots00\right\rangle
\ldots + \left|000 \ldots01\right\rangle \right)\]
diluted by white noise, 
if measurements in a Bell type experiment
 at \(N-2\) parties are made in the computational basis and all yield the \(-1\)
result (associated with $\left|0\right\rangle$), the remaining
pair of observers is left with a mixture of a \(2\)-qubit Bell
state with a substantially reduced amount of white noise. The
probability of such a chain of events  for the $N-2$ observers is
quite low. Nevertheless, we shall show that such an event contradicts,  very strongly
any local realistic description. In the case of the \(N\)-qubit
GHZ state diluted by white noise, scenario of this kind 
can  lead with unit probability to a 2-qubit Bell state without reduction of noise.
Because of that, the N-qubit states of the \(W\) class, for
$N>10$,  can lead to more robust, against noise admixture,
violations of local realism, than the $N$-qubit states of the GHZ
class. 
We also show that if an  \(N\)-qubit   \(W\) state and a GHZ state are separately 
sent through similar depolarizing channels,
and the 2-qubit state conditioned on measurements at the \(N-2\) parties is considered, 
the rate  of obtaining singlets (in specific distillation protocols) 
can be higher in the case of \(W\) state for a certain range of noise 
in the depolarizing channel. Importantly, in such scenarios, 
the \(W\) state performs better for \emph{higher} levels of noise, and this 
feature grows with \(N\).
These results may be of
importance  in quantum cryptography  and 
communication complexity \cite{GISIN,COMPLEXITY}. We also show
that these considerations lead to a relatively efficient
entanglement witness.

\section{Violation of local realism by W states}

To analyze how strongly  the \(N\)-qubit \(W\)
states violate local realism, we use 
the recently found multi-qubit correlation function Bell inequalities which form a necessary
and sufficient condition for the existence of a local realistic 
model for the correlation function in 
experiments with \emph{two}  local settings for each of the 
\(N\) observers \cite{WW, ZB}.

An \(N\)-qubit state \(\rho\) can always be written down as
\[\frac{1}{2^N} \sum_{x_1, \ldots, x_N = 0, x, y, z} T_{x_1 \ldots x_N}
\sigma_{x_1}^{(1)} \ldots \sigma_{x_N}^{(N)},\]
where \(\sigma_0^{(k)}\) is the identity operator and the
 \(\sigma_{x_i}^{(k)}\)'s (\(x_i = x, y, z\)) are the Pauli operators
of the \(k\)-th qubit. The coefficients
\[T_{x_1 \ldots x_N} = \tr(\rho\sigma_{x_1}^{(1)} 
\ldots \sigma_{x_N}^{(N)}), \quad (x_i = x, y, z)\]
are elements of the \(N\)-qubit correlation tensor
\(\hat{T}\) and they fully define the \(N\)-qubit correlation function \cite{ZB}.
A sufficient condition for the \(N\)-qubit correlation function
to satisfy all correlation function Bell inequalities  is that for \emph{any} set of
local coordinate systems, one must have \cite{ZB}
\begin{equation}
\label{suff}
\sum_{x_1, \ldots, x_N= x, y} T^2_{x_1 \ldots x_N} \leq 1,
\end{equation}
 the sum being taken over \emph{any} set of orthogonal pairs of axes of the local
coordinate systems of all observers.

Consider a mixture
 \(\rho^W_N\), of \(W_N\)
with white noise    \(\rho^{N}_{noise}=I^{(N)}/2^N\),  where
\(I^{(N)}\) is the unit operator in the tensor
product of  Hilbert spaces of the qubits:
\begin{equation}
\label{noisyWN}
\rho^W_N = p_N \left|W_N\right\rangle\left\langle W_N \right| +
(1 - p_N)\rho^{(N)}_{noise}.
\end{equation}
The parameter \(p_{N}\) will be called here ``visibility". It
defines to what extent the quantum processes associated with 
\(W_{N}\)  are ``visible" in those given by \(\rho_{N}^{W}\).
If the \(N\)-qubit correlations of
a pure state \(\left|\psi\right\rangle\)
are represented by a correlation tensor \(\hat{T}\), then the correlation tensor of a mixed
state \[p_{N}\left|\psi\right\rangle \left\langle\psi\right| +
(1-p_{N}) \rho^{(N)}_{noise} \] 
is  given by $\hat{T}^{'} = p_N \hat{T}$.

\subsection{The case of three qubits}

Consider now the mixture of the 3-qubit state \(\left|W_3\right\rangle\) (of visibility
\(p_3\)) with white noise. The correlation tensor of the pure state
\(\left|W_3\right\rangle\) is
\begin{equation}
\label{corrW3}
\begin{array}{llcl}
\hat{T}^{W_3}  =  \vec{z_1} \otimes \vec{z_2} \otimes \vec{z_3} \\
                         -  \frac{2}{3} \left(\vec{z_1} \otimes \vec{x_2} \otimes \vec{x_3}
+ \vec{x_1} \otimes \vec{z_2} \otimes \vec{x_3}  +     \vec{x_1} \otimes \vec{x_2} \otimes \vec{z_3}\right) \\
 -  \frac{2}{3} \left(\vec{z_1} \otimes \vec{y_2} \otimes \vec{y_3}  +  \vec{y_1} \otimes \vec{z_2} \otimes \vec{y_3} +
\vec{y_1} \otimes \vec{y_2} \otimes \vec{z_3}\right).
\end{array}
\end{equation}
where \( \vec{x_i}, \vec{y_i}, \vec{z_i}\) 
forms a Cartesian coordinate system with \(\vec{z_i}\) defining the computational basis.

Let us find the maximal value of the left hand side (lhs) of
(\ref{suff}) for  \(W_3\). We will show that in an
arbitrary set of local coordinate systems, \[\sum_{i,j,k = x, y}
T_{ijk}^2  \leq \frac{7}{3}.\] The value of the lhs of (\ref{suff}), for
the case when we are in the coordinate system in which
(\ref{corrW3}) is written down, is zero.   Moving into a
different  coordinate system can be always done for each observer
using three Euler rotations \cite{ZBML}. Therefore we shall assume that
first all three observers perform a rotation about the axes
\(\vec{z}_i\), then around the new \(\vec{x}_{i}{'}\) directions
and finally around \(\vec{z}_{i}{'}\) directions.

The set of first Euler rotations leaves the value of the lhs of
(\ref{suff}) at zero. The second rotation around
\(\vec{x}_i{'}\) axes leads to 
following values of the
components of the correlation tensor in the \(xy\) sector:
\(T^{''}_{xxx}\) is vanishing, whereas 
\[
\begin{array}{rcl}
T^{''}_{yyx} & = &  \frac{2}{3}   (\sin \phi_{23} \sin\theta_1 \cos\theta_2
              -        \sin \phi_{31} \cos\theta_1 \sin\theta_2),    \\
T^{''}_{yxy} & = &  \frac{2}{3}  (\sin \phi_{32} \sin\theta_1 \cos\theta_3
              -   \sin \phi_{21} \cos\theta_1 \sin\theta_3),        \\
T^{''}_{xyy} & = &  \frac{2}{3} (\sin \phi_{21} \sin\theta_3 \cos\theta_2
             - \sin \phi_{13} \cos\theta_3 \sin\theta_2),           \\
T^{''}_{xxy} & = &  \frac{2}{3} \cos \phi_{12} \sin\theta_3,  \quad
T^{''}_{xyx} =  \frac{2}{3} \cos \phi_{13} \sin\theta_2, \\
T^{''}_{yxx} & = &  \frac{2}{3} \cos \phi_{23} \sin\theta_1,
\end{array}
\]
(\(T^{''}_{yyy}\) is not explicitly needed)
where the \(\phi_i\)'s are the local angles of the first rotations and
\(\theta_i\)'s are those for the second one and \(\phi_{ij} =
\phi_i - \phi_j\). 
Employing   \( (A\cos \eta + B \sin\eta)^2 \leq
A^2 + B^2\) and that \( T^{''2}_{yyy} \leq 1\), one gets
\[
\begin{array}{ rcl}
 & \sum&_{i,j,k = x, y}
{T^{''2}_{ijk}}  \leq   1 + \frac{4}{9}(\sin^2\phi_{31} \cos^2 \theta_1 \\
  & + & \sin^2\phi_{23}\sin^2 \theta_1
  + \sin^2\phi_{21} \cos^2 \theta_1
 +  \sin^2\phi_{32} \sin^2 \theta_1 \\
 & + & \sin^2\phi_{13} \cos^2 \theta_3
  +  \sin^2\phi_{21} \sin^2 \theta_3
   + \cos^2\phi_{12} \sin^2 \theta_3 \\
 &  + & \cos^2\phi_{13} \sin^2 \theta_2
  + \sin^2 \phi_{23} \sin^2 \theta_1).
\end{array}
\]
The right hand side of this inequality is a linear function in \(\cos^2
\theta_i\)'s and \(\cos^2 \phi_{jk}\)'s. Thus its maximal value
is for  extreme values of these parameters, and 
\[\sum_{i,j,k = x, y} {T^{''2}_{ijk}}  \leq \frac{7}{3}.\] 
The last set of
Euler rotations around the \(\vec{z}_{i}{'}\) axes cannot change
the value of the lhs of this relation. So,  for any set of local
coordinate systems, one has \(\sum_{i,j,k = x, y} T_{ijk}^2
\leq 7/3.\) This inequality is saturated, as in the system
of coordinates in which (\ref{corrW3}) is written down,
\[\sum_{i,j,k = x, z} T_{ijk}^2  = \frac{7}{3}.\] 
(Note that \(y\) is
replaced by \(z\), in the last equation.)

For the noisy 3-qubit \(W\) state, one has \[\max\sum_{i,j,k = x, y}
T_{ijk}^2  = \frac{7}{3} p_3^2,\] and thus there is no violation of the
correlation function Bell inequalities  \cite{WW, ZB}
 for \[ p_3 \leq
\sqrt{\frac{3}{7}} \approx 0.654654.\] 
Note that at 
least one of the correlation function Bell 
inequalities is violated (as checked numerically) for \(p_3 \geq 0.65664\) \cite{LASKOWSKI}.

Surprisingly, if one takes a second look at the data, that can be
acquired in a three qubit correlation experiment, with the noisy
\(W_3\) state, one can lower the bound for \(p_3\) which allows
for a local realistic description (cf. \cite{Popescu}). Suppose that in the Bell
experiment, the observer \(3\) chooses as her/his measurements as
follows: the first observable is the  \(\sigma_z^{(3)}\)
operator, and the second one something else. The computational
basis for the third qubit is the
eigenbasis of \(\sigma_z^{(3)}\).
The measurements of \(\sigma_z^{(3)} = \left|1\right\rangle \left\langle 1\right| -
 \left|0\right\rangle \left\langle 0\right|\) will cause collapses of the full state 
\(\rho_3^{W}\) (cf. (\ref{noisyWN}))
into
new states. Whenever the result is \(-1\), the emerging state is of the following form:
$\rho^{W_{2}}_{(3)} \otimes \left|0\right\rangle \left\langle 0\right|$,
where the Werner state \(\rho_{(3)}^{W_{2}}\) reads
\begin{equation}
 \rho_{(3)}^{W_{2}} = p_{(3\rightarrow 2)} \left|W_2\right\rangle \left\langle W_2\right|
+ ( 1 - p_{(3\rightarrow 2)} ) \rho^{(2)}_{noise},
\end{equation}
with \( \left|W_2\right\rangle =  
(\left|01\right\rangle + \left|10\right\rangle)/\sqrt{2} \). The new
visibility parameter
 is given by
\begin{equation}
\label{recur23}
p_{(3\rightarrow 2)} = \frac{4p_3}{ 3 + p_3} \geq p_3.
\end{equation}
The 2-qubit state \(\rho_{(3)}^{W_{2}}\) has no local realistic description for 
\[1\geq p_{(3\rightarrow 2)}\geq \frac{1}{\sqrt{2}}.\] 
Therefore for
this range of \(p_{(3\rightarrow 2)}\), the results received by
the observers \(1\) and \(2\), which are conditioned on observer
\(3\) getting the \(-1 \) result (when s/he measures
\(\sigma_{z}^{(3)}\)), cannot have a local realistic model. We
have a subset of the data in the full experiment with no local
realistic interpretation. Surprisingly, the critical value of
\(p_3\) above which this phenomenon occurs is lower than
\(\sqrt{3/7}\):
\begin{equation}
\label{vis3}
p_3^{crit}= \frac{3}{4\sqrt{2} -1} \approx 0.644212 < \sqrt{\frac{3}{7}} < 0.65664,
\end{equation}
and can be obtained by putting \(p_{(3\rightarrow 2)} = 1/\sqrt{2}\) in (\ref{recur23}).

The correlation functions are averages of
products of the local results, and as such do not distinguish the
situation when, e.g., local results are \(+1, +1, -1\) (for the
respective observers), with the one when the results are \(+1,
-1, +1\). Therefore, an analysis of the results of the first
 two parties conditioned on the third party receiving e.g. \(-1\), can lead to a
more stringent constraint on local realism. And this is exactly
what we have received here. Note that no sequential measurements
are involved (the third party performs just one measurement).
Communication between the parties is  only after the experiment,
just to collect the data.

For the \(W_3\) state, this refinement of data analysis seems
to be  optimal. To test the ultimate critical value for \(p_3\),
we  employed the numerical procedure based on linear
optimization, which tests whether the full set of
\emph{probabilities} involved in a \(N\) particle experiment
admits an underlying local realistic model. The procedure has
been described in many works \cite{KASZL}, and therefore will not
be given here. Since the program analyzes the full set of
\emph{probabilities}, its verdict 
is based on the \emph{full} set
of data available in the Bell experiment. The program 
has found that Bell type experiments on a noisy \(W_3\) state have always a
local realistic description for \(p_3\) below the numerical
threshold of \(0.644212\).
 That is, we have a full agreement, up to the numerical accuracy, with the result given
in eq. (\ref{vis3}).

\subsection{ The case of N qubits}

For the \( N>3\), the above phenomenon gets even more pronounced.
The sufficient condition for the local realistic description
(\ref{suff}), can be shown to be satisfied for
\begin{equation}
\label{NWWWZB}
p_N \leq \sqrt{\frac{N}{3N - 2}}.
\end{equation}
The method that we have used to get (\ref{NWWWZB}) is the straightforward generalization of the Euler
rotations method to the \(N\) qubit case. Since
\[ \lim_{N\rightarrow \infty} \sqrt{\frac{N}{3N - 2}} = \frac{1}{\sqrt{3}},\]
explicit local realistic
models for the \(N\) qubit correlation functions (in a standard Bell experiment, involving pairs
of alternative observables at each site) exists for large \(N\) for \(p_3\) as high as
\(1/\sqrt{3}\). We have also numerically found the threshold value of \(p_N\),
above which at least one of the correlation function Bell 
 inequalities is violated. It does not differ too much from the right hand side of
(\ref{NWWWZB})
 and e.g. for \(N=4\), it reads \(0.63408\) and for \(N=11\), it is \(0.59897\).

\begin{figure}[tbp]
\vspace{1cm}
  \epsfig{figure=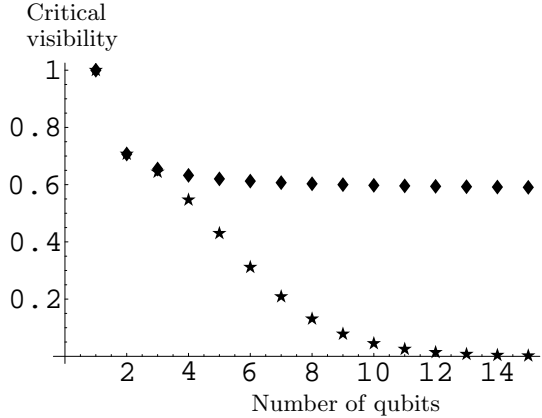,width=0.40\textwidth}
\put(-110,-8){Number of qubits}
\put(-195,140){Critical}
\put(-195,130){visibility}
\caption{The stars are a plot of critical visibility (a higher visibility gives violation of local 
realism), \(p_{N}^{crit}\), 
 obtained
by the method of projections and the diamonds are a plot of the
values \(\sqrt{N/(3N - 2)}\) of \(p_N\),  below which exist
 a local hidden variable description of the \(N\)-qubit
correlation functions,  for the noisy \(W\) state of \(N\)
qubits.} \label{crit}
\end{figure}

However the more refined method of data analysis leads to different results. Imagine
 now that the last \(N-2\) observers have the \(\sigma_z\) observable within their local pair of
alternative observables in the
Bell test. Then if all of them get the \(-1\) result, the collapsed state will be given by
\[ \rho_{(N)}^{W_2} \otimes \left(\otimes_{i=3}^{N} \left|0\right\rangle_{ii}\left\langle 0 \right|\right), \]
with the 2-qubit state \(\rho_{(N)}^{W_2}\) being
\begin{equation}
\label{eq_projected}
\rho_{(N)}^{W_2} = p_{(N\rightarrow 2)} \left|W_{2}\right\rangle \left\langle W_{2}\right| +
(1- p_{ (N\rightarrow 2)}) \rho^{(2)}_{noise}
\end{equation}
where
\begin{equation}
\label{N|2a} p_{(N\rightarrow 2)} = \frac{p_N}{p_N+ (1 -
p_N)\frac{ N}{2^{N-1} }}.
\end{equation}
Since for \(p_{(N\rightarrow 2)} > 1/ \sqrt{2}\), no local realistic description of this subset of data is
possible,
the critical visibility \(p_N\), which does not allow a local realistic model reads
\begin{equation}
\label{N|2}
p_{N}^{crit} = \frac{N}{(\sqrt{2}-1) 2^{N-1} + N}.
\end{equation}
Note that \(p_N^{crit} \rightarrow 0\) when \(N\rightarrow \infty\).
For sufficiently large $N$
the decrease has an exponential character! This behavior is strickingly different than the one for the
threshold value
of \(p_N\) which is sufficient to satisfy the \(N\) qubit correlation function Bell  inequalities (Fig. \ref{crit}).

For low \(N\), the value of 
\(p_N^{crit} \) was confirmed by the
numerical
procedure \cite{KASZL} mentioned earlier ( which analyzes the full set of data for the problem). The critical numerical values
 are \(p_4^{thr} = 0.546918\), which is exactly equal to the
\(6\)-digit approximation of the \(p_4^{crit}\) value given by
(\ref{N|2}), and
 \(p_5^{thr} = 0.4300 \), which
is equal to the \(4\)-digit approximation of \(p_5^{crit}\) of
(\ref{N|2}).
 This suggests a
conjecture that (\ref{N|2}) is the real threshold. However if this is untrue, the decrease in \(p_N\)
must be even bigger!

\section{Comparison of GHZ and W states from the perspective of violation of  local realism}
\label{section-comparison-lhv}

The  GHZ states exhibit maximal violations of the correlation function Bell inequalities  \cite{WW, ZB}, and the
 violations
 when measured by the threshold visibility \(p_N^{GHZ}\) exhibit an exponential behavior. Let us now
compare these two families.
A noisy \(N\) particle GHZ state, given by
\[ \rho_N^{GHZ} = p_N
^{'} \left|GHZ\right\rangle \left\langle GHZ \right| + (1-
p_N^{'}) \rho^{(N)}_{noise}\]  
violates correlation function Bell  inequalities
whenever \[p_N ^{'}> p_N^{GHZ} = \frac{1}{\sqrt{2^{N-1}}}.\] We have
performed for \(N=3,4,5\) the numerical analysis of the
possibility of the existence of a local realistic model for the full set of data for 
GHZ correlations, and the returned numerical critical values
fully agree with this value. For \(N \geq 11\), \( p_N^{crit}\)
of (\ref{N|2}) for the \(W\) states is lower than the one for GHZ
ones. The \(11\) or more qubit \(W\) state violate local realism
more strongly than their GHZ counterparts, and this increases
exponentially.

\subsection{Violation of local realism using functional Bell inequalities}

It may seem  that the GHZ states will regain their glory
of being the most nonclassical ones, if one introduces more
alternative measurements, than just two, for each observer.
However let us note that there exists a sequence of functional
Bell inequalities \cite{Zf} for \(N\) qubits, which involve the
entire range of
local measurements in one plane. We will now show that even if we consider these 
functional Bell inequalities, the \(W\) states remain more nonclassical than the GHZ states. 
But before that, we first briefly discuss the functional Bell inequalities.

\subsubsection{The functional Bell inequalities}

The functional Bell inequalities \cite{Zf}
 essentially follow from
 a simple geometric observation that
 in any real vector space, if for two vectors \(h\) and \(q\) one has
 \(\left\langle h \mid q\right\rangle < \parallel q \parallel^2\),
 then this
 immediately implies that \(h \ne q\). In simple words, if the
 scalar product of two vectors has a lower value than the length
of one of them, then the two vectors cannot be equal.

 Let \(\varrho_N\) be a state shared between \(N\) separated parties.
 Let \(O_n\) be an arbitrary observable at the \(n\)th location (\(n=1,\ldots,
 N\)).
 The quantum mechanical prediction \(E_{QM}\) for the
 correlation in the state \(\varrho_N\), when
 these observables are measured, is
 \begin{equation}
 \label{EQM}
 E_{QM}\left(\xi_1, \ldots, \xi_N\right) = \tr \left(O_1 \ldots O_N \varrho_N\right),
 \end{equation}
 where \(\xi_n\) is the aggregate of the local parameters at the \(n\)th site.
 Our object is to see whether this prediction can be reproduced
 in a local hidden variable theory. A local hidden variable correlation
 in the present scenario must be of the form
 \begin{equation}
 \label{EHV}
 E_{LHV}\left(\xi_1, \ldots, \xi_N\right) = \int d\lambda \rho (\lambda) \Pi_{n=1}
 ^{N}
 I_{n} ( \xi_{n}, \lambda),
 \end{equation}
 where \(\rho(\lambda)\) is the distribution of the local hidden variables and
 \( I_{n}(\xi_{n}, \lambda) \) is the predetermined measurement-result of the
 observable
 \(O_n(\xi_n)\) corresponding to the hidden variable \(\lambda\).

 Consider now the scalar product
 \begin{equation}
 \begin{array}{lcl}
 \left\langle E_{QM}\mid E_{LHV}\right\rangle = \\
 \int  d\xi_1 \ldots
 d\xi_N 
  E_{QM}\left(\xi_1, \ldots, \xi_N\right)
 E_{LHV}\left(\xi_1, \ldots, \xi_N\right)
 \end{array}
 \label{EQMEHV}
 \end{equation}
 and the norm
 \begin{equation}
 \label{normEQM}
 \parallel E_{QM} \parallel^2 =
 \int  d\xi_1 \ldots d\xi_N
 \left(E_{QM}\left(\xi_1, \ldots, \xi_N\right)\right)^{2}.
 \end{equation}
 If we can prove that a strict inequality holds, namely
 for all possible \(E_{LHV}\), one has 
\begin{equation}
\label{star}
\left\langle E_{QM}\mid E_{LHV} \right\rangle \leq B,
\end{equation}
with the number \(B < \parallel E_{QM} \parallel^2\), we will immediately have
 \(E_{QM} \ne E_{LHV}\), indicating that the correlations in
 the state \(\varrho_N\) are of a different character than in
 any local realistic theory. We then could say that the state \(\varrho_N\) violates
 the ``functional" Bell inequality (\ref{star}), as this 
Bell inequality is expressed in terms
 of a typical scalar product for square integrable functions. Note that the 
value of the product depends on a
 continuous
range of parameters (of the measuring apparatuses) at each site.

\subsubsection{Comparison of \(W\) state with GHZ state when 
the latter violates functional Bell inequalities}

The critical visibility for which the GHZ state violates the functional Bell inequalities 
\cite{Zf} is 
lower than that for which it violates the multiqubit two-settings Bell inequalities \cite{WW,ZB}.
Precisely, the critical
visibility above which a functional Bell inequality is violated by an 
\(N\)-qubit GHZ state, is
\(2 \left(2/\pi\right)^{N}\) \cite{Zf,ZK-func}. 
This is obtained for measurement settings in the \(x-y\)
planes for all observers sharing the GHZ state.
This critical visibility is better than that
for violation of the multiqubit Bell inequalities for \(N \geq 4\). 
Nevertheless, the \(W\) family leads to
stronger violations of local realism for \(N \geq 15\) (compare
with eq. (\ref{N|2})).
Interestingly, the \(W\) states
do not violate the functional Bell inequalities
involving all settings in one plane for \(N>3\).

\section{The case of \(G\) states}

It may seem that the results obtained in section \ref{section-comparison-lhv} 
depend  on the fact that the GHZ state has
 an equal number of \(\left|0 \right\rangle\)s  and \(\left|1\right\rangle\)s, when expressed  in the 
\(\sigma_z\) basis, 
while the \(W\)-state has an asymmetry in this respect. In this section, we show that this is not the case.

Consider  for example the \(N\) (\(\geq 3\)) qubit state \cite{amader_crypto}
\[ \left|G_N\right\rangle =  \frac{1}{\sqrt{2}}(\left|W_N\right\rangle  + \left|\overline{W}_N\right\rangle), \] 
where \(\left |\overline{W}_N\right\rangle \)  is obtained  by interchanging 
\(\left|0\right\rangle\) and \(\left|1\right\rangle\) in \(\left|W_N\right\rangle  \). 
For \(N=2\), we define \(\left|G_2\right\rangle = \left|W_2\right\rangle = \left|\overline{W}_2\right\rangle\).
This state (\(\left |G_N\right\rangle \)) has an equal number of \(\left|0\right\rangle\)s and 
\(\left|1\right\rangle\)s, just 
as in the GHZ states.   Now consider the state \(G_N\) admixed with white noise,
\[\rho^{G}_N= q_N \left|G_N\right\rangle \left\langle G_N\right| + ( 1 - q_N) \rho^{N}_{noise}.\]
In a similar process as described before, if  \(N-2\) parties make  measurements in the \(\sigma_z\) basis 
and when all of them obtain
\(+1\) or all of them obtain \(-1\), the state obtained at the remaining two parties is the (two-qubit)
Werner state (similarly as in eq. (\ref{eq_projected}))  
\[
\rho_{(N)}^{G_2} = q_{(N\rightarrow 2)} \left|G_{2}\right\rangle \left\langle G_{2}\right| +
(1- q_{ (N\rightarrow 2)}) \rho^{(2)}_{noise}
\]
 with
\[q_{ (N \rightarrow 2)} =  \frac{q_N}{q_N + (1-q_N) \frac{N}{ 2^{N-2}}}.\]
We can then proceed just as we did 
in section \ref{section-comparison-lhv}, in
the case of \(W\) state.

The state \(\rho_{(N)}^{G_2}\) has no  
local realistic description  for 
\(q_{(N\rightarrow 2)} > 1/\sqrt{2}\) and this  implies that the state \(\rho^{G}_N\) cannnot 
have a local realistic model for
\[ q_{N} > q_{N}^{crit} \equiv  \frac{N}{N + (\sqrt{2}-1)2^{N-2}}.\]
For \(N \geq 13\), 
\[q_N^{crit} \leq p_N^{GHZ}.\]
Therefore for \(N \geq 13\), the \(G\) states also violate local realism more strongly than GHZ states. 
 However the nonclassicality in the G-states is less pronounced than that in the W-states. For the \(W\) states, the
cross-over was at \(N =11\) (see 
 section \ref{section-comparison-lhv}).

\section{Comparison between GHZ and W states with respect to yield of singlets}
\label{section-yield}

The method of data analysis presented above implies another difference
between the \(W\) states and the GHZ states. For a noisy GHZ state,
if one of the observers performs a measurement in the basis
\( \left\{(\left|0\right\rangle \pm  \left|1\right\rangle)/\sqrt{2}\right\} \),
 the projected state,
has for the other \(N-1\) observers again the form of  a noisy
GHZ state, and the visibility parameter is the \emph{same} as
before. Further, after \(N-2\) observers perform  measurements
in this basis, whatever are their results, the last two observers
share a noisy Bell state, with the same visibility as the
original noisy GHZ state.  In contrast, noisy
(\(N\)-qubit)  \(W\) states, upon a measurement of $\sigma_z$,
 by the first observer, resulting in $-1$, lead  to a noisy \(W\) state \(\rho_{N-1}^{W}\),
 for
the remaining \(N-1\) observers, of \emph{increased visibility}, namely to
\[p' \left|W_{N-1}\right\rangle \left\langle W_{N-1}\right|
+ ( 1- p') \rho_{noise}^{(N)}\] with
\[ p' \equiv p_{(N \rightarrow N-1)} = \frac{p_N}{p_N+ (1-p_N)\frac{N}{2  (N-1)}} \geq p_N.\]
 The other result, $+1$, leads  to a separable  mixture of
$\left|00 \ldots 0\right\rangle $ with  white noise.

If one's aim is to get  a Bell state in the hands of just two observers
of the required (high) visibility, say at least \(p_2^{R}\), then this can always
be achieved with some probability by taking a noisy \(W\) state of sufficiently many
qubits, and performing \(N-2\) measurements of \(\sigma_z\) on \(N-2\) qubits.
 The success is conditioned
on all results being \(-1\). For the given \(p_2^{R}\), the noisy \(W\) state
of visibility \(p_N\)
must be for the number of qubits, \(N\), for which
\[ p_2^{R} > \frac{p_N}{p_N+ (1-p_N)\frac{N}{2^{N-1}}}.\]
Therefore if one's aim is to have a Bell state
between two observers with as high visibility as possible, one can send
through a noisy channel a \(W_N\) state. From the noisy \(W_N\) state, in a probabilistic way,
one can extract a high visibility Bell state.
No such possibility exists for the GHZ state.

Consider now the situations when 
a large number of copies of the \(N\)-qubit \(W\) state and  GHZ state are separately 
sent through a depolarizing channel of visibility \(p\) \cite{channel}. For  \(W_N\), 
using the method of projections, 
one 
is able to obtain with probability \[\mathbb{P} = \frac{2p}{N + (1-p)/2^{N-2}},\] a Bell state 
of visibility (see eq. (\ref{N|2a})) 
\begin{equation}
\label{lagey-tuk}
v=\frac{p}{p+ (1-p) \frac{N}{2^{N-1}}}.
\end{equation}  
For the state \(GHZ_N\), one obtains 
with unit probability a Bell state with the original visibility \(p\). 
We will now compare the yield of singlets in these two situations, by using (in both cases) two different 
protocols for distilling the resulting two-qubit state, obtained after the specific projections on the multiqubit 
states.

\subsection{One-way hashing protocol for distillation}

\begin{figure}[tbp]
\vspace{1cm}
  \epsfig{figure=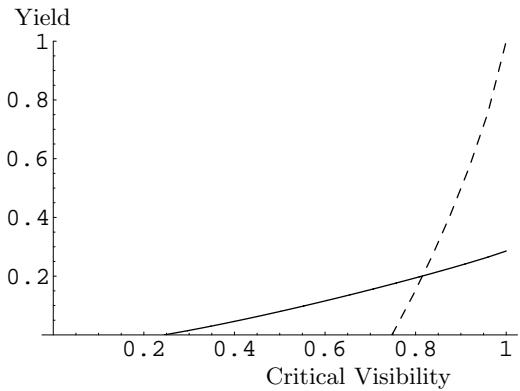,width=0.40\textwidth}
\put(-100,-8){Critical Visibility}
\put(-195,128){Yield}
\caption{Plot of visibility in the depolarizing channel
vs. the yield in the one-way hashing protocol. 
The undashed line is for the state \(W_{7}\) while the dashed line is for \(GHZ_{7}\).} \label{yield}
\end{figure}

Using the one-way hashing protocol
for distillation \cite{huge}, the per copy yield of singlets (of arbitrarily high fidelity) is 
\[(1 - S(\varrho(v)))\mathbb{P}\] 
for the state \(W_N\) and \(1-S(\varrho(p))\) for the state \(GHZ_N\), where 
\[\varrho(x) = x\left|W_2\right\rangle \left\langle W_2 \right| + (1-x)\rho^{(2)}_{noise}\]
 and \[S(\eta) = -\tr(\eta \log_2 \eta)\] is the 
von Neumann entropy of \(\eta\). As shown in Fig. \ref{yield} (for \(N = 7\)), 
the yield for the \(W\) state 
is better for a large range of \(p\) than that for the GHZ state. 
This feature remains for all \(N\) and gets pronounced with increasing \(N\). 
Even for \(N=3\), there is a small range of \(p\), in which the yield of singlets
is higher for the \(W\) state.
And importantly, the ranges in which the \(W\) states are better are for \emph{higher} levels of noise.

\subsection{Two-way recurrence-hashing distillation protocol}

Similar features are obtained for \emph{two-way} distillable 
entanglement. For example, although  one-way distillable
entanglement is vanishing for \(\varrho(1/2)\), the two-way 
recurrence-hashing distillation protocol gives a positive yield \cite{huge}.
To get \(\varrho(1/2)\) from, say \(W_7\), the visibility in the depolarizing channel 
can be as low as 
\[0.098592.\] 
(This value is obtained by putting \(v = 1/2\) and \(N=7\) in eq. (\ref{lagey-tuk}).) 
For the same visibility in the channel,  \(GHZ_7\) produces the 
\emph{separable} state \(\varrho(0.098592)\) \cite{money-achhey-to}. 
We therefore obtain a \emph{qualitative} difference between the \(W\) state and the GHZ state in this respect.

\section{A simple entanglement witness}

Looking at the projected state can also serve as a simple
entanglement witness. The state \[\rho = \varepsilon \rho{'} +
\frac{(1 - \varepsilon)}{2^N} I_{2^N},\] where \(\rho{'}\) is a
normalized density matrix, is separable for \cite{ball} 
\[\varepsilon < \frac{1}{1+ \frac{2^{N}}{2}}.\]  Choosing \(\rho{'}\) as the
\(N\)-qubit \(W\) state and using eq. (\ref{N|2a}),
 one obtains
 an upper bound \[\frac{1}{1+ \frac{2^{N}}{N}}\] of the radius
\(\varepsilon\) of the separable ball (as a
noisy Bell state is separable for
$p_{(N\rightarrow 2)}\leq 1/3$ \cite{money-achhey-to}), which is of the same order
as obtained in Ref. \cite{ball}.

It is interesting whether states of other families have similar
surprising properties, which can emerge after specific
measurements by  some of the observers.

\begin{acknowledgements}
We acknowledge  discussions with Wies{\l}aw Laskowski.
A.S. and  U.S. acknowledge  University of Gda\'{n}sk,
 Grant No. BW/5400-5-0256-3 and  EU grant EQUIP,
No. IST-1999-11053. M.Z. is supported  by Professorial subsidy of the Foundation for Polish Science.
\end{acknowledgements}


\begin{thebibliography}{10}
\bibitem{Acin1} A related result was obtained in Ref. \cite{Acin}, where
it was shown that a non-maximally entangled state in two qutrits violates 
a Bell inequality (derived in \cite{CGLMP}) 
more strongly than any other state. This Bell inequality was shown 
to be necessary and sufficient for correlation functions describing experiments 
with three local settings at each site 
\cite{Masanes}.
\bibitem{channel}A depolarizing channel of visibility \(p\), produces \[p\varrho + (1-p)\rho_{noise}^{(N)},\]
for an \(N\)-qubit state \(\varrho\) sent through it.
\bibitem{money-achhey-to}The state 
\(\varrho(x) = x\left|W_2\right\rangle \left\langle W_2 \right| + (1-x)\rho^{(2)}_{noise}\) is separable 
for \(x \leq 1/3\) \cite{huge}.
\end{thebibliography}
\end{document}